\documentclass[twocolumn]{aastex7}

\begin{document}

\title{A Definitive Determination of the Interstellar Carbon Abundance toward $\rho$ Ophiuchi A and B}

\correspondingauthor{Adam M.~Ritchey}

\author[orcid=0000-0002-3659-4192]{Adam M.~Ritchey}
\affiliation{Department of Physics and Astronomy, University of Toledo, Toledo, OH 43606, USA}
\email[show]{ritchey.astro@gmail.com}  

\author[orcid=0000-0002-8433-9663]{S.~R.~Federman}
\affiliation{Department of Physics and Astronomy, University of Toledo, Toledo, OH 43606, USA}
\email{steven.federman@utoledo.edu}

\author[orcid=0000-0001-6277-9556]{Daniel~E.~Welty}
\affiliation{Space Telescope Science Institute, Baltimore, MD 21218, USA}
\email{dwelty@stsci.edu}

\author[orcid=0000-0003-0760-4483]{Adolf~N.~Witt}
\affiliation{Department of Physics and Astronomy, University of Toledo, Toledo, OH 43606, USA}
\email{adolf.witt@utoledo.edu}

\begin{abstract}
We present the results of an effort to derive interstellar gas-phase \ion{C}{2} abundances along the lines of sight toward $\rho$~Oph A and B. Our analysis is based on high-resolution NUV and FUV archival spectra acquired with the Space Telescope Imaging Spectrograph on the Hubble Space Telescope. Column densities of \ion{C}{2} are derived both from fits to the weak \ion{C}{2}]~$\lambda2325$ intersystem transition and from fits to the damping wings of the strong \ion{C}{2}~$\lambda1334$ line. We find that the results from the weak-line and strong-line determinations agree with each other remarkably well for both sight lines, demonstrating the reliability of the $f$-values of the \ion{C}{2} transitions. Furthermore, the gas-phase C abundance that we obtain for $\rho$~Oph A is in very good agreement with previous determinations of interstellar C abundances from measurements of the weak \ion{C}{2}]~$\lambda2325$ line. By demonstrating the reliability of the damping wing fitting technique for the \ion{C}{2}~$\lambda1334$ line, our analysis opens the door to future surveys of interstellar C abundances using the same methodology.
\end{abstract}


\section{Introduction}
The gas-phase interstellar carbon abundance is a crucial parameter that bears on the ionization balance in neutral diffuse interstellar clouds and on the abundance of carbonaceous dust grains in the interstellar medium (ISM). Determinations of interstellar \ion{C}{2} abundances (C$^+$ being the dominant ionization state of C in diffuse clouds) have typically followed one of three approaches. The most secure determinations are derived from observations of the weak \ion{C}{2}]~$\lambda2325$ intersystem transition in absorption against background stars \citep[][]{c96,s04}. This line is optically thin along diffuse and translucent sight lines so that the column density is directly proportional to the equivalent width. However, the line is typically extremely weak (with equivalent widths of $\sim$1$-$2 m\AA{} or less) and therefore requires high resolution, very high signal-to-noise (S/N) ratio NUV spectra to be detected. Another method for deriving \ion{C}{2} abundances involves fitting the damping wings of the strong, typically heavily saturated \ion{C}{2}~$\lambda1334$ transition \citep[][]{s11,p12}. However, this approach can yield unreliable results if the underlying continuum is poorly defined or if there are low column density \ion{C}{2} absorption components at moderately high velocity that obscure the Lorentzian damping wings of the main low-velocity interstellar \ion{C}{2} component. A third approach involves examining absorption or emission from the forbidden [\ion{C}{2}] 158 $\mu$m line \citep[][]{d97,g15,c25}. In this case, uncertainties arise from the assumptions being made about the excitation conditions in the interstellar gas and from attempts to disentangle absorption and emission along complicated lines of sight.

Prior to 2011, most measurements of interstellar C abundances were derived using the weak intersystem transition at 2325.4 \AA{}. With the exception of a few outliers, these earlier measurements are generally consistent with an average gas-phase carbon abundance of C/H $\sim$ $1.6\times10^{-4}$ \citep[e.g.,][]{s04}. However, the total number of sight lines with secure detections of the weak \ion{C}{2}]~$\lambda 2325$ feature is small ($\sim$13) and the prospects for expanding the sample are limited by the need for high-resolution, very high S/N ratio NUV spectra, which generally require long exposure times with the only instrument currently capable of obtaining such data, the Space Telescope Imaging Spectrograph (STIS) on the Hubble Space Telescope (HST). Thus, \citet{s11} explored using the strong \ion{C}{2}~$\lambda1334$ transition to derive C abundances via fits to the Lorentzian damping wings. \citet{s11} reported \ion{C}{2} column densities (using their strong-line method) for the same six sight lines that \citet{s04} had analyzed via the weak intersystem line. Unfortunately, the results did not agree, with the strong-line determinations yielding column densities that were $\sim$40\% smaller than those from the weak-line measurements. Based on these results, \citet{s11} suggested that the oscillator strength of the intersystem transition was in need of revision.

In this Letter, we revisit the comparison between column densities of \ion{C}{2} derived from fits to the damping wings of the allowed \ion{C}{2}~$\lambda1334$ transition and from profile fits to the weak, semi-forbidden \ion{C}{2}]~$\lambda2325$ line.\footnote{Hereafter, we will drop the half bracket notation and refer to the intersystem transition simply as \ion{C}{2}~$\lambda2325$.} Our analysis is based on high-resolution FUV and NUV HST/STIS spectra of $\rho$~Oph A (HD~147933) and $\rho$~Oph B (HD~147934). Ultimately, we show that there is no tension between the column densities obtained from the strong and weak \ion{C}{2} transitions, the results are consistent with earlier determinations of interstellar C abundances, and the oscillator strengths ($f$-values) of the \ion{C}{2} transitions are secure.

\section{Observations and Data Processing}
HST/STIS observations of $\rho$~Oph A and B were obtained in 2015 under program GO 13365 (PI: Martin Cordiner). The observations employed both the E140H grating, with the wavelength setting at 1271 \AA{}, and the E230H grating, with the wavelength setting at 2263 \AA{}, allowing coverage of both \ion{C}{2} transitions at high-resolution ($R\sim120,000$). The sight lines to $\rho$~Oph A and B were chosen for this study because (1) the component structure along the two lines of sight is relatively simple, with $\sim$90\% of the column density in the `main' low-velocity component (as determined from the weak \ion{O}{1}~$\lambda1355$ line), (2) the total hydrogen column density is quite large ($10^{21.7}$ cm$^{-2}$ toward $\rho$~Oph A), meaning that the weak \ion{C}{2}~$\lambda2325$ line is easily detectable toward both $\rho$~Oph A and $\rho$~Oph B, and (3) the stars are part of the same binary system, which means we can compare column densities from the strong and weak \ion{C}{2} transitions for two closely spaced sight lines as an additional check on the accuracy of our results.

The pipeline-calibrated STIS spectra were retrieved from the Mikulski Archive for Space Telescopes (MAST). An IDL procedure was used to coadd multiple exposures of a given target obtained with the same echelle grating and to combine adjacent echelle orders into a single FUV or NUV spectrum for each sight line.\footnote{The IDL procedure (stack\_echelle\_fits) was written by Edward B.~Jenkins.} When combining multiple overlapping spectra, the procedure weights the individual flux values by the inverse square of the associated uncertainties. It is particularly important to properly combine the overlapping portions of the echelle orders when analyzing the strong \ion{C}{2}~$\lambda1334$ line, which typically appears in two adjacent orders of the E140H grating. As described below, our analysis of the \ion{C}{2}~$\lambda2325$ feature utilizes the \ion{O}{1}~$\lambda1355$ line as a profile template. The \ion{O}{1}~$\lambda1355$ line also appears in two adjacent orders of the E140H grating, which are near the edge of the spectral format when the wavelength setting at 1271 \AA{} is chosen. The standard CALSTIS pipeline procedure extracts only one of the orders containing \ion{O}{1}~$\lambda1355$. Thus, to improve the overall S/N ratio in this line, we manually extracted both orders containing \ion{O}{1}~$\lambda1355$ from the calibrated 2D spectral images obtained from MAST, following the procedure described in \citet{r11}, and then coadded the orders to produce final \ion{O}{1}~$\lambda1355$ spectra.

\section{Column Density Determinations}
\subsection{Comparison of Oscillator Strengths}
As an explanation of the disparate results on \ion{C}{2} column densities derived from the strong and weak transitions, \citet{s11} suggested that the oscillator strength of the weak intersystem line may not be accurate. However, this suggestion is not supported by a review of \ion{C}{2} oscillator strengths reported in the literature. \citet{m03} adopts a value of $\log f\lambda=-3.954\pm0.003$ for the \ion{C}{2}~$\lambda2325$ transition. This value (and its small uncertainty) is derived from the experimental results of \citet{t99}, who measured precise intercombination transition rates using a heavy-ion storage ring. Other experimental and theoretical determinations \citep[e.g.,][]{s99,tff00,ch02} agree with the oscillator strength from \citet{t99} at the 5\% level (see Table~\ref{tab:fvalues}). Thus, in light of the good agreement among the various experimental and theoretical determinations, the $f$-value of the \ion{C}{2}~$\lambda2325$ transition appears to be well established.

\begin{deluxetable*}{lcccc}
\tablecolumns{5}
\tablecaption{Comparison of Oscillator Strengths for \ion{C}{2} Transitions\label{tab:fvalues}}
\tablehead{ \colhead{Transition} & \colhead{$\lambda_0$} & \colhead{$\log f\lambda$} & \colhead{Error} & \colhead{Reference} \\
\colhead{} & \colhead{(\AA{})} & \colhead{} & \colhead{(dex)} & \colhead{} }
\startdata
$2s^2 2p~^2{P}^{\rm o}_{1/2}-2s2p^2~^2{D}_{3/2}$ & 1334.532 & 2.227 & \ldots & 1 \\
 & & 2.233 & \ldots & 2 \\
 & & 2.236 & \ldots & 3 \\
$2s^2 2p~^2{P}^{\rm o}_{1/2}-2s2p^2~^4{P}_{1/2}$ & 2325.403 & $-3.940$ & 0.018 & 4 \\
 & & $-3.954$ & 0.003 & 5 \\
 & & $-3.935$ & \ldots & 6 \\
 & & $-3.952$ & \ldots & 7 \\
\enddata
\tablerefs{(1) \citet{ns81}, (2) \citet{y87}, (3) \citet{fft04}, (4) \citet{s99}, (5) \citet{t99}, (6) \citet{tff00}, (7) \citet{ch02}.}
\end{deluxetable*}

As a dipole-allowed transition, the strong \ion{C}{2}~$\lambda1334$ line is generally regarded as having a well-determined oscillator strength. The close-coupling calculations of \citet{y87} yield a value of $\log f\lambda=2.233$ for the $\lambda1334$ transition. \citep[This is essentially the value adopted in][]{m03}. Various other theoretical determinations \citep[e.g.,][]{ns81,fft04} agree with this result at the 1\% level (Table~\ref{tab:fvalues}). For the analysis described below, the $f$-values reported in \citet{m03} are adopted for both \ion{C}{2} transitions.

\subsection{Weak-line Measurements}
Our analysis of the \ion{C}{2}~$\lambda2325$ features toward $\rho$~Oph A and $\rho$~Oph B followed the methodology we have adopted in previous analyses of weak interstellar absorption lines \citep{r11,r18}. Small segments of spectra ($\sim$2 \AA{} wide) centered on the \ion{C}{2}~$\lambda2325$ (and \ion{O}{1}~$\lambda1355$) lines were isolated and then continuum-normalized by fitting low-order polynomials to regions free of interstellar absorption. The S/N ratio (per pixel) near the \ion{C}{2}~$\lambda2325$ lines is $\sim$90 for both sight lines. As anticipated, the \ion{C}{2}~$\lambda2325$ (and \ion{O}{1}~$\lambda1355$) absorption profiles are nearly identical for the two closely spaced sight lines.\footnote{The similarity in the absorption profiles of various atomic and molecular species toward $\rho$~Oph A and B can also be seen in the optical spectra published by \citet{p04}.} We find equivalent widths of $W_{\lambda}=1.9\pm0.5$ m\AA{} for the weak \ion{C}{2} line in both directions. The \ion{O}{1}~$\lambda1355$ equivalent widths are $15.8\pm0.2$ m\AA{} for $\rho$~Oph A and $16.2\pm0.3$ m\AA{} for $\rho$~Oph B.

\begin{figure*}
\centering
\includegraphics[width=0.49\textwidth]{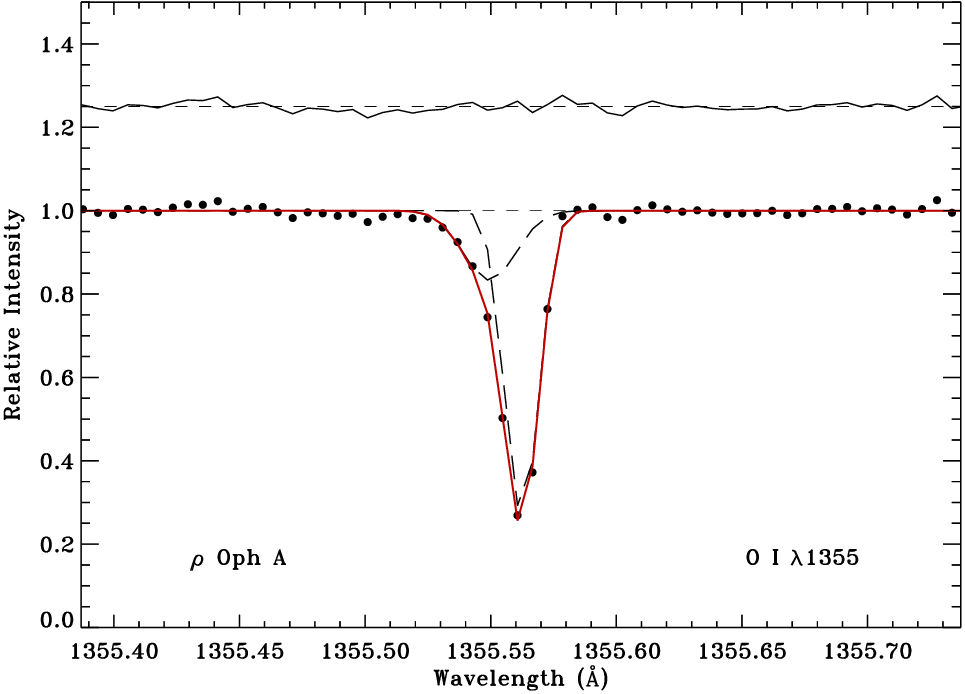}
\includegraphics[width=0.49\textwidth]{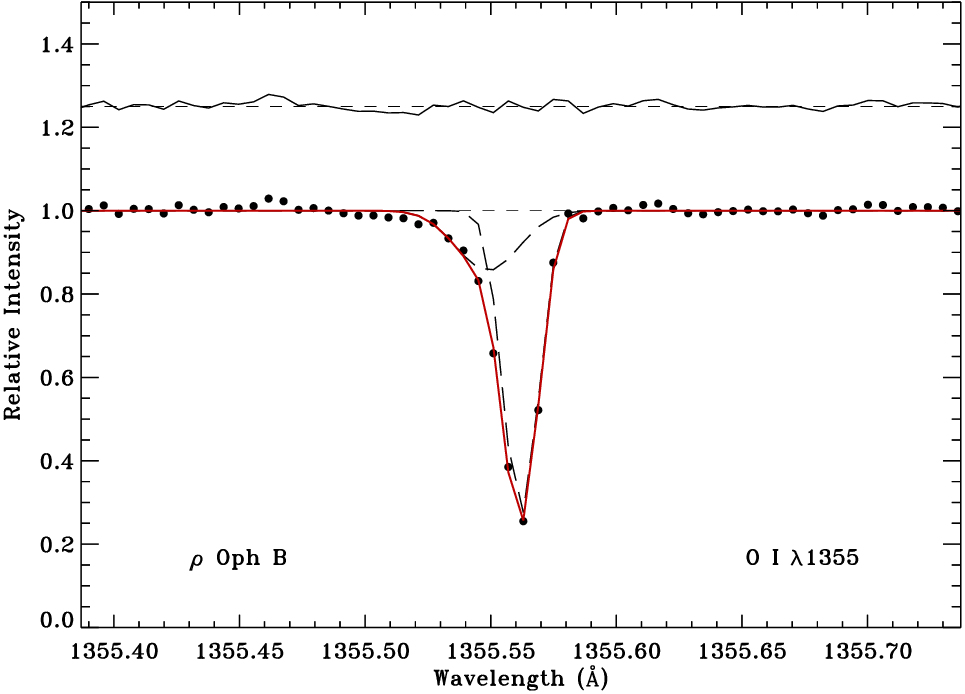}
\includegraphics[width=0.49\textwidth]{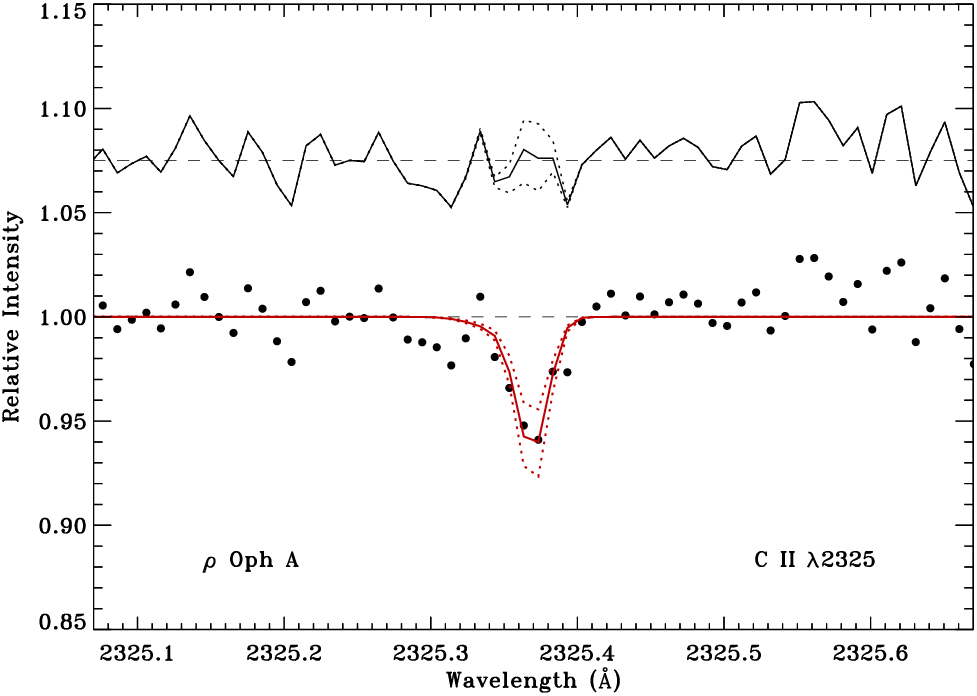}
\includegraphics[width=0.49\textwidth]{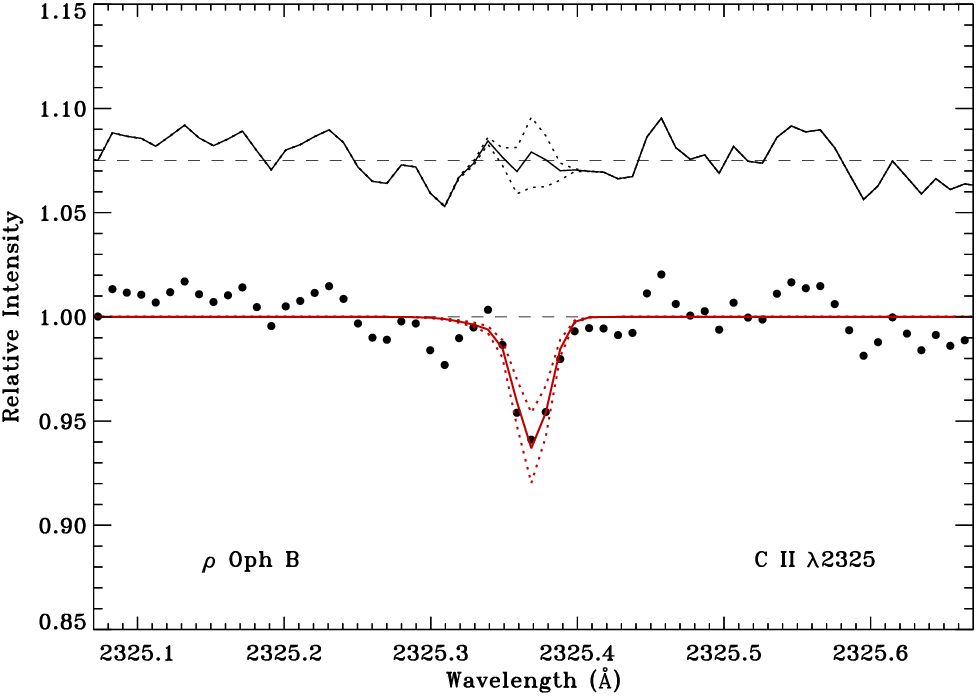}
\caption{Profile synthesis fits to the \ion{O}{1}~$\lambda1355$ and \ion{C}{2}~$\lambda2325$ lines toward $\rho$~Oph A (left panels) and $\rho$~Oph B (right panels). The data points represent the observed spectra, while the synthetic profiles are shown by the solid red lines. The dashed curves in the upper panels indicate the results for the two individual velocity components obtained from the fits to the \ion{O}{1} lines. The dotted red curves in the lower panels correspond to the $\pm1\sigma$ range in the derived \ion{C}{2} column densities. The fit residuals are plotted above each spectrum. The same range in velocity is shown in each panel.\label{fig:weak_fit}}
\end{figure*}

Despite the relatively high S/N of the NUV STIS spectra, the weakness of the \ion{C}{2} absorption features makes it difficult to discern the detailed component structure in the line profiles. Thus, we use the stronger intersystem transition of \ion{O}{1} at 1355.6 \AA{} to construct a profile template for \ion{C}{2}~$\lambda2325$. Since both O$^0$ and C$^+$ are dominant ions, and are typically only lightly depleted into interstellar dust grains \citep[e.g.,][]{c04,s04}, the absorption profiles of \ion{O}{1} and \ion{C}{2} are expected to be similar. Our profile fitting methodology makes use of the code ISMOD \citep{s08}, which derives the best-fitting column densities, velocities, and $b$-values for a specified number of components by minimizing the rms deviations in the fit residuals. An initial examination of the \ion{O}{1}~$\lambda1355$ lines toward $\rho$~Oph A and B revealed a single dominant component at (a heliocentric velocity of) $-7.9$ km~s$^{-1}$ and a much weaker component at $-10.8$ km~s$^{-1}$. We therefore fit the \ion{O}{1}~$\lambda1355$ lines with two components finding total column densities of $\log N($\ion{O}{1}$)=18.28\pm0.06$ toward $\rho$~Oph A and $18.26\pm0.05$ toward $\rho$~Oph B. For both directions, $\sim$86\% of the \ion{O}{1} column density is associated with the main component, which has a $b$-value of $\sim$1.1 km~s$^{-1}$. The minor component is somewhat broader with an average $b$-value of 3.1 km~s$^{-1}$. Our profile synthesis fits to the \ion{O}{1}~$\lambda1355$ lines toward $\rho$~Oph A and B are shown in Figure~\ref{fig:weak_fit}.

The component structures derived from the fits to the \ion{O}{1}~$\lambda1355$ lines were used as fixed profile templates when fitting the \ion{C}{2}~$\lambda2325$ features. In particular, the fractional column densities, relative velocities, and $b$-values of the two line-of-sight components were held fixed in ISMOD fits to the \ion{C}{2}~$\lambda2325$ lines. Only the total \ion{C}{2} column density and a velocity offset for the profile as a whole were permitted to vary. The resulting profile synthesis fits for the \ion{C}{2}~$\lambda2325$ lines are presented in Figure~\ref{fig:weak_fit}. The derived total \ion{C}{2} column densities are given in Table~\ref{tab:col_den}. The uncertainties in the column densities derived from the weak \ion{C}{2} transition are entirely dominated by the statistical fluctuations in the measured intensities. The rms fluctuations in the (normalized) continuum regions near the \ion{C}{2}~$\lambda2325$ line are $\sim$0.011 for both directions. We multiply the rms in the continuum by the FWHM of each fitted component to obtain the uncertainty in the equivalent width. The uncertainties for the individual components are then added in quadrature to yield the uncertainty in the total equivalent width. Since the \ion{C}{2}~$\lambda2325$ line is on the linear portion of the curve of growth, the percent error in equivalent width is adopted as the percent error in column density. These correspond to the errors shown in Table~\ref{tab:col_den}. Figure~\ref{fig:weak_fit} illustrates how the synthetic profile (and the fit residuals) would vary according to the $\pm1\sigma$ uncertainties in the derived \ion{C}{2} column density.

\begin{deluxetable}{lccc}
\tablecolumns{4}
\tablecaption{Derived \ion{C}{2} Column Densities\label{tab:col_den}}
\tablehead{ \colhead{Star} & \colhead{$\log N$(1334)} & \colhead{$\log N$(2325)} & \colhead{$\log N$(\ion{C}{2})\tablenotemark{a}} }
\startdata
$\rho$~Oph A & $17.947\pm0.064$ & $17.941\pm0.106$ & $17.946\pm0.056$ \\
$\rho$~Oph B & $17.937\pm0.082$ & $17.942\pm0.107$ & $17.939\pm0.067$ \\
\enddata
\tablenotetext{a}{Weighted mean of the results from the \ion{C}{2}~$\lambda1334$ and $\lambda2325$ lines.}
\end{deluxetable}

\begin{figure*}
\centering
\includegraphics[width=0.49\textwidth]{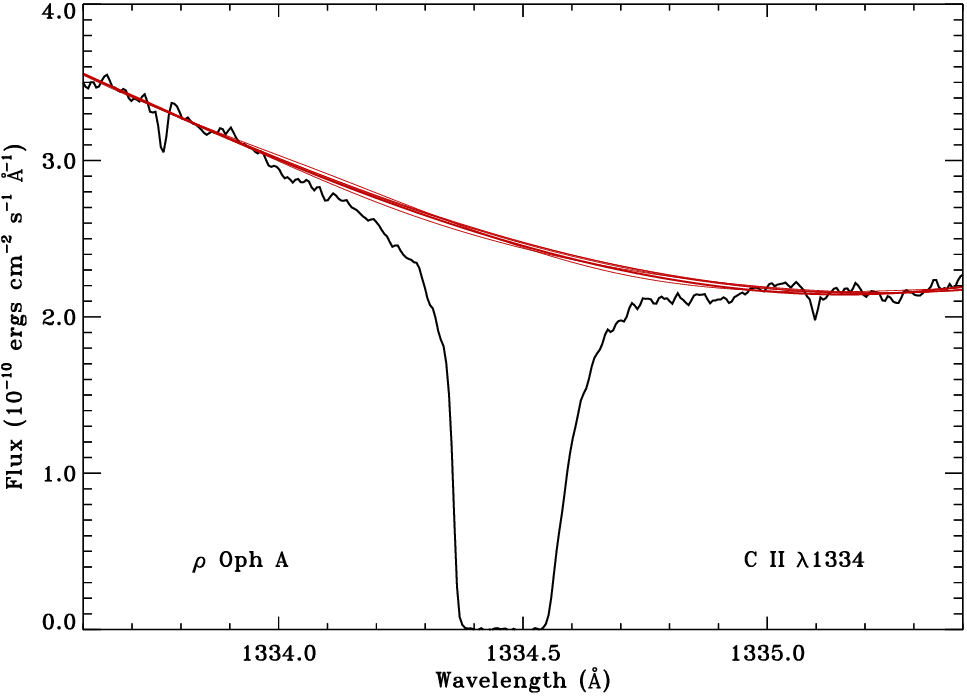}
\includegraphics[width=0.49\textwidth]{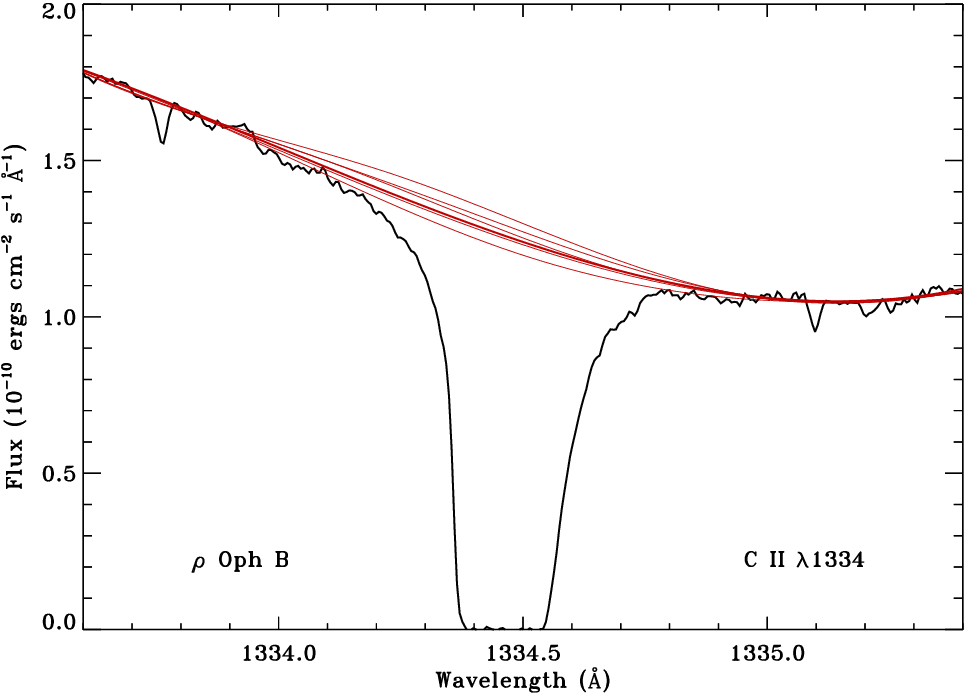}
\caption{High-resolution HST/STIS spectra in the vicinity of the interstellar \ion{C}{2}~$\lambda1334$ line toward $\rho$~Oph A (left panel) and $\rho$~Oph B (right panel). The thick red line in each panel indicates the adopted continuum fit. The thinner red lines demonstrate the range of plausible continuum fits used to determine the effect of continuum placement on the derived column density. (The two weak and narrow absorption features on either side of the \ion{C}{2} line are due to vibrationally-excited H$_2$.)\label{fig:strong_cont}}
\end{figure*}

\subsection{Strong-line Determinations}
We chose the sight lines to $\rho$~Oph A and B for our analysis, not only because the weak \ion{C}{2}~$\lambda2325$ lines are unambiguously detected, but also because the Lorentzian damping wings in the strong \ion{C}{2}~$\lambda1334$ transition are clearly visible in both directions. A major uncertainty in deriving \ion{C}{2} column densities from fits to the damping wings of the $\lambda1334$ transition is the shape of the underlying continuum onto which the interstellar absorption features are superimposed. In the case of $\rho$~Oph A and B (which have spectral types B1 V and B2 V), the interstellar \ion{C}{2}~$\lambda1334$ lines (along with the nearby interstellar \ion{C}{2}*~$\lambda1335$ features) are superimposed onto very broad stellar \ion{C}{2} absorption lines. (The FWHM of the stellar \ion{C}{2} line is $\sim$500 km~s$^{-1}$ for both stars.) Previous efforts to derive \ion{C}{2} abundances from fits to the damping wings \citep{s11,p12} employed synthetic stellar models to fit the continuum near the interstellar \ion{C}{2} lines. While such an approach may be appealing from a theoretical standpoint, the problem is that the synthetic spectrum invariably does not provide a perfect match to the actual observed spectrum. The continuum fitting procedure adopted in \citet{s11}, for example, required seven free parameters to fine-tune the modelled continuum, and even then there are cases where the synthetic spectrum does not appear to match the observed spectrum particularly well.

For our analysis, we adopt a more direct approach to fitting the continuum. First, we isolate the interstellar \ion{C}{2}~$\lambda1334$ line within a small spectral window with a width of 2 \AA{}. (A width of 2 \AA{} at 1334.5 \AA{} corresponds to a velocity width of 450 km~s$^{-1}$.) We then fit a low-order polynomial to regions very far from the line and well outside the obvious damping wings. (Essentially, we choose continuum points at the very edges of the spectral window.) For both sight lines, we found that a first-order cubic spline function provided a good representation of the underlying continuum. Our continuum fits for the \ion{C}{2}~$\lambda1334$ lines toward $\rho$~Oph A and B are shown in Figure~\ref{fig:strong_cont}. We also explored how changes in the order of the polynomial, and differences in the placement of the continuum points, affected the derived column densities. These alternate continuum fits are also shown in Figure~\ref{fig:strong_cont}. For $\rho$~Oph A, the alterations in the fitting parameters did not significantly change the resulting continuum fit. For $\rho$~Oph B, however, subtle differences in the structure of the noise in the continuum regions led to more significant variations in the alternate continuum fits.

\begin{figure*}
\centering
\includegraphics[width=0.49\textwidth]{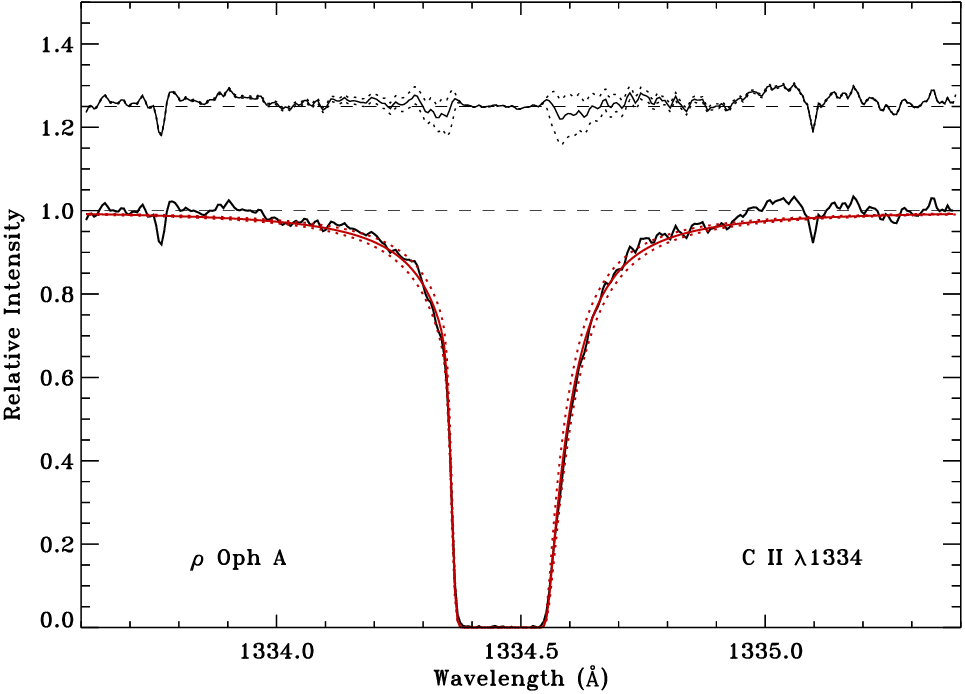}
\includegraphics[width=0.49\textwidth]{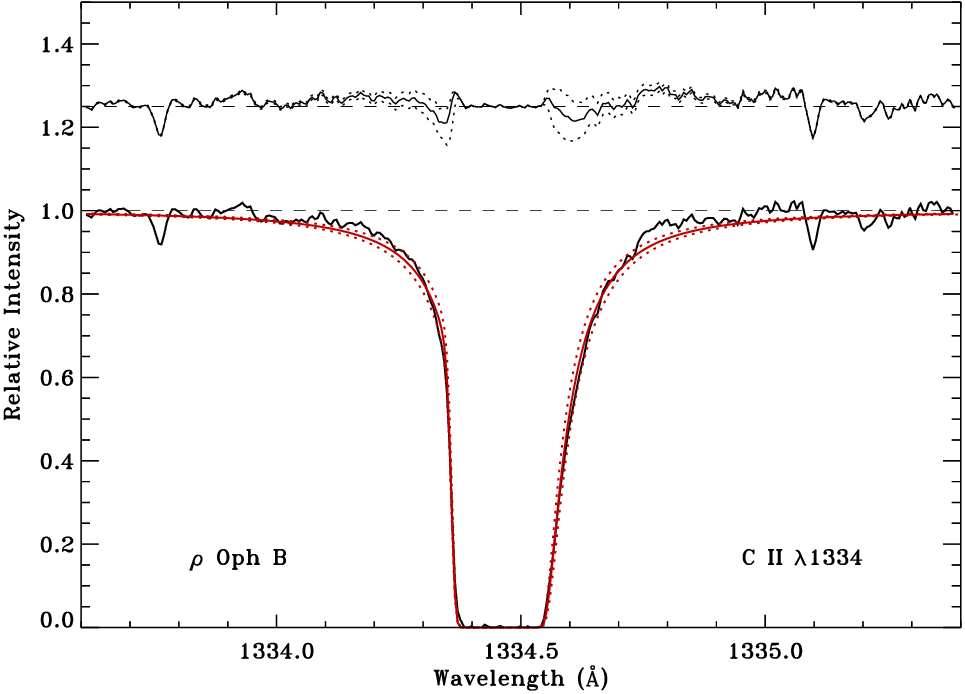}
\caption{Normalized spectra of the interstellar \ion{C}{2}~$\lambda1334$ line toward $\rho$~Oph A (left panel) and $\rho$~Oph B (right panel). The solid red line in each panel indicates the best-fitting Voigt profile. The dotted red lines correspond to the $\pm1\sigma$ range in the derived column density due to statistical uncertainties. The fit residuals are plotted above each spectrum.\label{fig:strong_fit}}
\end{figure*}

Once the data were continuum-normalized, we used our multi-component Voigt profile fitting routine ISMOD to fit the \ion{C}{2}~$\lambda1334$ absorption lines.\footnote{We make no attempt to fit the \ion{C}{2}*~$\lambda1335$ absorption features. The column density of \ion{C}{2} in the excited fine-structure level is generally too small to create significant damping wings (i.e., the lines are on the flat portion of the curve of growth). Any attempt to derive column densities from these features would be accompanied by exceptionally large uncertainties.} For the $\rho$~Oph A and B sight lines, the damping wings are dominated by a single low-velocity absorption component (i.e., the `main' component seen in the \ion{O}{1}~$\lambda1355$ profile). However, excess absorption on the blue side of the \ion{C}{2} profile requires that additional low column density components be included in the fit. The component structure cannot be directly evaluated using the heavily saturated \ion{C}{2} profile. Thus, we used a combination of the \ion{O}{1}~$\lambda1355$ line, which probes the highest column density components, and the \ion{S}{2}~$\lambda1250$ line, which is useful for examining the relative strengths of the lower column density components at higher velocity. An initial prediction for the component structure in the \ion{C}{2} line was obtained from the results of ISMOD fits to the \ion{O}{1} and \ion{S}{2} profiles (adjusting for differences in the cosmic abundances of the elements and in the expected amount of dust depletion).

Ultimately, we fit the \ion{C}{2}~$\lambda1334$ absorption profiles with six components, the two prominent components seen in the \ion{O}{1}~$\lambda1355$ line, and four additional components at $-30.7$ km~s$^{-1}$, $-26.8$ km~s$^{-1}$, $-18.8$ km~s$^{-1}$, and +0.0 km~s$^{-1}$. In these fits, we held the relative velocities among the components fixed, but all other parameters were permitted to vary. We stress that the total \ion{C}{2} column density obtained with this method is almost entirely determined by the strength of the main absorption component. The inclusion of the additional velocity components is necessary in order to properly fit the observed spectrum. However, the contribution from those components to the total \ion{C}{2} column density is minimal (i.e., $\sim$13\%, which is less than the associated uncertainties). Since the component structure cannot be determined directly from the saturated \ion{C}{2} profile, the individual column densities of those minor additional components are not well constrained in the fit. However, this has very little impact on the outcome for the total \ion{C}{2} column density.

Our Voigt profile fits to the \ion{C}{2}~$\lambda1334$ lines toward $\rho$~Oph A and B are presented in Figure~\ref{fig:strong_fit}. The total \ion{C}{2} column densities derived from these fits are given in Table~\ref{tab:col_den}. There are two major sources of uncertainty in the column densities derived from the strong \ion{C}{2} transition. First, there are the statistical uncertainties in the measured intensities that affect the Voigt profile fit. The quality of the fit may be judged by the degree to which the damping wings from the synthetic absorption profile match the observed spectrum. To examine the effect of statistical uncertainties, we increased and decreased the total column density of the synthetic profile to match the fluctuations in the intensity points in the wings of the profile. In particular, the wing regions closest to the core of the line are the most important in the fitting process. Small changes in the synthetic profile in these regions lead to large variations in the fit residuals. For both sight lines, we found that the allowed range in column density due to statistical uncertainties corresponded to $\pm0.06$ dex. (This range in column density is indicated by the dotted red lines in Figure~\ref{fig:strong_fit}.)

In addition to the statistical uncertainties, there are systematic effects related to the placement of the continuum. This is particularly important for strong-line determinations where the strength of the damping wings can be greatly impacted by the choice of continuum. For each of the alternate continuum fits described above (see Figure~\ref{fig:strong_cont}), we employed the same Voigt profile fitting procedure as for our nominal continuum fit to derive the total \ion{C}{2} column density. We then evaluated the standard deviation of the resulting values. For $\rho$~Oph A, the range of plausible continuum fits was small and the resulting spread in the derived column densities was $\pm0.02$ dex. For $\rho$~Oph B, the variation in continuum fits was larger and the spread in column densities corresponded to $\pm0.06$ dex. These systematic uncertainties were added in quadrature with the statistical uncertainties to arrive at the values given in Table~\ref{tab:col_den}.

\section{Discussion and Conclusions}
The main conclusion from our analysis is that the \ion{C}{2} column densities derived from the weak \ion{C}{2}~$\lambda2325$ line and the strong \ion{C}{2}~$\lambda1334$ feature agree with each other remarkably well for both sight lines toward $\rho$~Oph A and B (Table~\ref{tab:col_den}). Furthermore, the column densities we find toward $\rho$~Oph A are indistinguishable from those toward $\rho$~Oph B, as would be expected given the binary nature of the targets and the similarity in the absorption profiles. We also note that the uncertainties associated with the strong-line determinations are smaller than those from the weak-line measurements, indicating that, in cases where the damping wings are clearly visible, Voigt profile fitting of the \ion{C}{2}~$\lambda1334$ feature can be a reliable method for deriving \ion{C}{2} abundances.

Taking the weighted mean of the results from the strong-line and weak-line determinations, we find total \ion{C}{2} column densities of $\log N($\ion{C}{2}$)=17.95\pm0.06$ for $\rho$~Oph A and $17.94\pm0.07$ for $\rho$~Oph B.\footnote{While it is difficult to determine a precise column density for \ion{C}{2}* from the available spectra, the lack of significant damping wings in the \ion{C}{2}* $\lambda1335$ line suggests that the contribution from \ion{C}{2}* to the total \ion{C}{2} column density is $\sim$6\% or less for both sight lines. This would correspond to a logarithmic increase in $\log N($\ion{C}{2}$)$ of $\lesssim0.03$ dex.} The atomic and molecular hydrogen column densities toward $\rho$~Oph A are $\log N($\ion{H}{1}$)=21.63\pm0.09$ \citep{ds94} and  $\log N($H$_2$$)=20.57\pm0.15$ \citep{b78}, yielding a total hydrogen column density of $\log N($H$_{\rm tot}$$)=21.70\pm0.08$. While published column densities of \ion{H}{1} and H$_2$ are not available for $\rho$~Oph B, presumably they would be very similar to those toward $\rho$~Oph A. Indeed, our own analysis of the \ion{H}{1} Ly$\alpha$ absorption features from the available STIS spectra yield column densities of $\log N($\ion{H}{1}$)=21.60\pm0.03$ for $\rho$~Oph A and $21.61\pm0.03$ for $\rho$~Oph B. These results are consistent not only with each other but also with the \ion{H}{1} column density from \cite{ds94}. There is no direct measurement of $N$(H$_2$) toward $\rho$~Oph B. However, the similarity in the column density of CH (and CN) toward $\rho$~Oph A and B \citep{p04} suggests that the H$_2$ column density would also be similar in the two directions.

Adopting the previously published column densities of \ion{H}{1} and H$_2$ toward $\rho$~Oph A, and our derived value for the \ion{C}{2} column density, we find a gas-phase carbon abundance of C/H = $(1.77\pm0.46)\times10^{-4}$. While some of the C in diffuse clouds exists in the form of C$^0$ and CO, these species do not contribute meaningfully to the total gas-phase C abundance toward $\rho$~Oph A. \citet{z97} find a total \ion{C}{1} column density toward $\rho$~Oph A of $\log N($\ion{C}{1}$)=15.52\pm0.01$, while \citet{f03} report a CO column density of $\log N($CO$)=15.28\pm0.05$. The combined contributions from \ion{C}{1} and CO are thus 0.6\% of the total C column density (which corresponds to a logarithmic increase of 0.003 dex). The dust-phase C abundance toward $\rho$~Oph A may be obtained by subtracting the gas-phase abundance from an assumed total interstellar abundance. \citet{l25} report a proto-solar C abundance of C/H = $(3.98\pm0.92)\times10^{-4}$. If this represents the present-day total interstellar (gas + dust) abundance of C, then the implied dust-phase C abundance toward $\rho$~Oph A would be $221\pm102$ parts per million of H (ppm).

The gas-phase C abundance we derive toward $\rho$~Oph A is entirely consistent with previous determinations of C abundances derived from analyses of the weak intersystem transition \citep[e.g.,][and references therein]{s04}. In particular, the weighted mean of 14 previous determinations, which includes one measurement based on fitting the damping wings of the strong \ion{C}{2} transition toward 23 Ori \citep{w99}, is $<$C/H$>$ = $(1.70\pm0.14)\times10^{-4}$. However, our results are not consistent with more recent attempts to obtain \ion{C}{2} abundances from the strong $\lambda1334$ transition \citep{s11,p12}. \citet{s11} obtained \ion{C}{2} column densities that were consistently below those derived from the weak transition measured along the same lines of sight, leading those authors to suggest that the $f$-value of the intersystem line was in need of revision. We find nearly identical \ion{C}{2} column densities from analyses of the $\lambda1334$ and $\lambda2325$ transitions toward $\rho$~Oph A and B, demonstrating that there is no tension between the $f$-values of these transitions. Moreover, as shown in Table~\ref{tab:fvalues}, there is consensus among theoretical and experimental efforts that the oscillator strengths of the \ion{C}{2} transitions are secure.

The \ion{C}{2} column densities derived by \citet{p12} using the $\lambda1334$ transition span a much larger range than those of most earlier efforts. Their determinations of gas-phase C abundances, which incorporate the results of \citet{s11}, span the range from C/H = 69 ppm to 464 ppm, with an average of 201 ppm and a standard deviation of 116 ppm. The likely reason for the large spread in derived C abundances is the considerable systematic uncertainties involved in obtaining \ion{C}{2} column densities from the $\lambda1334$ transition. We chose to study $\rho$~Oph A and B because the line-of-sight component structure is very simple and the damping wings in the $\lambda1334$ line are easy to identify and to fit. Many sight lines are not so well-suited for this type of analysis. In many directions, there are weaker \ion{C}{2} absorption components at higher velocities that tend to obscure the damping wings from the main low-velocity component. This appears to be the case for many of the sight lines analyzed by \citet[][see their Figure 2]{p12}. While fits to the damping wings of the \ion{C}{2}~$\lambda1334$ line can be a reliable method for obtaining gas-phase C abundances, the sight lines must be carefully chosen to avoid these uncertainties.

In summary, we examined high-resolution NUV and FUV HST/STIS spectra for the sight lines toward $\rho$~Oph A and B in an effort to derive \ion{C}{2} abundances from both the weak \ion{C}{2}~$\lambda2325$ intersystem transition and the strongly saturated \ion{C}{2}~$\lambda1334$ line. Our main findings are that the \ion{C}{2} column densities derived from the two transitions agree with each other remarkably well for both sight lines, demonstrating that the oscillator strengths of the two \ion{C}{2} transitions are reliable and well established. The C abundance obtained for the line of sight to $\rho$~Oph A is consistent with many previous determinations of C abundances derived from the weak intersystem transition, providing further clarity on the gas-phase abundance of C in the ISM. By demonstrating the reliability of the damping wing fitting technique for analyses of the \ion{C}{2}~$\lambda1334$ line, our analysis opens the door to future surveys of interstellar C abundances using the same methodology, with the caveat that the sample of sight lines must be carefully selected to avoid the systematic uncertainties that seem to have plagued previous investigations.

\begin{acknowledgments}
This research is based on observations made with the NASA/ESA Hubble Space Telescope obtained from the Space Telescope Science Institute, which is operated by the Association of Universities for Research in Astronomy, Inc., under NASA contract NAS 5-26555. The specific observations are associated with program GO 13365 (PI: Martin Cordiner) and can be obtained from the MAST archive at \dataset[doi:10.17909/n0xy-tc24]{\doi{10.17909/n0xy-tc24}}.
\end{acknowledgments}


\facilities{HST(STIS)}

\software{ISMOD \citep{s08}}

\end{document}